\documentclass[aip,cp,amsmath,amssymb,reprint]{revtex4-2}

\usepackage{graphicx}% Include figure files
\usepackage{dcolumn}% Align table columns on decimal point
\usepackage{bm}% bold math
\usepackage[mathlines]{lineno}% Enable numbering of text and display math
\usepackage[utf8]{inputenc}
\usepackage[T1]{fontenc}
\usepackage{hyperref}
\usepackage{bm}% bold math   
\usepackage{braket}
\usepackage{slashed}
\usepackage{multirow}
\usepackage{epsfig}
\usepackage{epstopdf}
\usepackage{bbm}
\usepackage{mathptmx}

\def\la{\lambda} \def\lap{\lambda^{\prime}}  \def\de{\delta}  \def\dag{\dagger}
    \def\S{{\rm S}}
\def\bnabla{{\bm \nabla}}

\def\nn{\nonumber}

\begin{document}

\title{Spin Structure of heavy-quark hybrids}

\author{Jaume Tarr\'us Castell\`a} 
 \email[]{jtarrus@ifae.es}
\affiliation{ Grup de F\'isica Te\`orica, Dept. F\'isica and IFAE-BIST, Universitat Aut\`onoma de Barcelona, E-08193 Bellaterra (Barcelona)}

\date{\today}
\begin{abstract}
Exotic quarkonia are candidates to new types of hadrons including four quarks or gluonic degrees of freedom as constituents, the latter being a unique feature of QCD. We review recent developments in nonrelativistic EFTs to describe exotic quarkonia and in particular recent results on the spectrum of heavy hybrids including spin-dependent contributions up to $1/m_Q^2$-terms in the heavy-quark-mass expansion. We determine the nonperturbative contributions to the matching coefficients of the EFT by fitting our results to lattice-QCD determinations of the charmonium hybrid spectrum and extrapolate the results to the bottomonium hybrid sector where lattice-QCD determinations are still challenging. We also report on a recent new approach to quarkonium hadronic transitions that does not use the twist expansion and uses the hybrid spectrum as the intermediate octet states.
\end{abstract}

\maketitle

\section{Introduction}

In the past decade many new, unexpected states, that do not fit the standard quarkonium picture, have been found in the quarkonium spectrum close and above open flavor thresholds. These states are commonly referred as exotic quarkonium. Exotic quarkonia are extremely interesting because they are candidates for nontraditional hadronic states, that is states that cannot be classified as mesons or baryons. Two new types of hadron have been proposed: states formed by a heavy-quark pair and a gluonic excitation and states formed by a heavy-quark and a light-quark pair. These states are usually referred as (heavy) hybrid (quarkonium) and tetraquarks respectively. In the latter case several pictures emerge depending on the spatial arrangement of the heavy and light quarks, from compact tetraquarks to shallow bound states of heavy-light mesons. Furthermore, it has also been proposed that some of these states have been in fact misidentified and their signals are due to kinematic effects.

Although many different models have been proposed to study exotic quarkonium the only two approaches connected to the underlying theory of the strong interactions, QCD, are effective field theories (EFT) and lattice QCD. These two approaches are in fact complimentary. Lattice QCD can provide \textit{ab initio} results from QCD, however when dealing with a system with well separated scales the computational costs can be difficult to overcome. On the other hand, EFTs make use of precisely these energy gaps to provide a description the exotic quarkonium systems, but requires of lattice input in order to determine the nonperturbative matching coefficients. Our aim is to build such EFTs and provide specific predictions for the spectrum, decays and transitions.

Exotic quarkonia are formed by two distinct components, the heavy-quark antiquark pair and the light degrees of freedom, which can be gluonic or light-quark in nature. The primary characteristic of exotic quarkonia is that the heavy quarks mass is larger than the typical hadronic scale $m_Q\gg\Lambda_{\rm QCD}$. Therefore the heavy quarks are nonrelativistic and the exotic states can be studied using NRQCD. The second important characteristic is that the typical binding energy of the heavy quarks is smaller than the energy scale of the light degrees of freedom states $\Lambda_{\rm QCD}\gg E_b$. This has lead to the observation that exotic quarkonium can be studied in an adiabatic expansion between the dynamics of its two components with similarities to the Born-Oppenheimer approximation in diatomic molecules ~\cite{Griffiths:1983ah,Juge:1997nc,Braaten:2014qka,Brambilla:2017uyf}. 

In diatomic molecules the nuclei pair are the heavy degrees of freedom playing a role analogous to the heavy quarks, while the electrons  are the light degrees of freedom analogous to the gluons and light-quarks in exotic quarkonium. In the Born-Oppenheimer approximation the electronic energy levels are computed in the limit of static nuclei and are called electronic static energies. The vibrational energy levels associated to the two-nuclei dynamics are formed around the minima of the static energies. The separation between the electronic static energies is of order $m_e\alpha^2$, with $m_e$ the electron mass and $\alpha$ the fine structure constant, while the separation between vibrational levels is of order $m_e\alpha^2\sqrt{m_e/m_N}$, with $m_N$ the nuclei mass. Thus, we observe the same separation between the energy scales of light and heavy degrees of freedom also present in exotic quarkonium.

\section{\label{stcen}Static energies}

The natural point to start the study of exotic quarkonium is the determination of the static energies. These are defined as the energies of the eigenstates of a $Q\bar{Q}$ pair in NRQCD in the static limit. The static energies are characterized by the following quantum numbers: $\bm{r}$ the separation between quark and antiquark, the light-quark flavor (for simplicity we will only consider isospin) and the representation $\Lambda^{\sigma}_{\eta}$ of $D_{\infty\,h}$. According to this symmetry, the static eigenstates are classified in terms of the absolute value of the angular momentum along the heavy quark-antiquark axis ($|\lambda|\equiv\Lambda=0,\,1,\,2\dots$, to which one gives the traditional names $\Sigma$, $\Pi$, $\Delta\dots$), $CP$ ($g$ for even or $u$ for odd), and the reflection properties with respect to a plane that passes through the quark-antiquark axis (+ for even or - for odd). Only the $\Sigma$ states are not degenerate with respect to the reflection symmetry.

The specific form of these static eigenstates depends on nonperturbative physics and are unknown, nevertheless the corresponding energy eigenvalues, the static energies, can be obtained from large time logarithms of appropriate correlators

\begin{align}
E^{(0)}_{n}(r)=\lim_{T\to\infty}\frac{i}{T}\log \langle \mathcal{O}_n(T,\,\bm{r},\,\bm{R})|\mathcal{O}_n(0,\,\bm{r},\,\bm{R})\rangle\,,\label{ste}
\end{align}
where $n$ stands for the set of quantum numbers that identify the static eigenstate, $\bm{R}$ and $\bm{r}$ are the center of mass and relative coordinates of the heavy-quark pair and $\mathcal{O}_n$ is an interpolating operator.

For hybrid and tetraquark states an appropriate interpolating operator reads

\begin{align}
\mathcal{O}_n(t,\,\bm{r},\,\bm{R})=\chi(t,\,\bm{R}-\bm{r}/2)\phi(t,\,\bm{R}-\bm{r}/2,\bm{R})H_n(t,\,\bm{R})\phi(t,\,\bm{R},\bm{R}+\bm{r}/2)\psi^\dagger(t,\,\bm{R}+\bm{r}/2)\,,\label{intop}
\end{align}
with $H_n(t,\,\bm{R})$ a gluonic operator or light-quark operator from table~\ref{gop}, $\psi$ the Pauli spinor field that annihilates a quark, $\chi$ the one that creates an antiquark and $\phi$ is a Wilson line. The correlator in Eq.~\eqref{ste} with the interpolating operator of Eq.~\eqref{intop} corresponds to a static Wilson loops with the insertion in the spatial sides of the $H_n$ light degree of freedom operator.

\begin{table}[ht!]
\centering
\caption{Examples of gluonic operators and light-quark operators for quarkonium hybrids and tetraquarks respectively, $\bm{q}=(u,\,d)$ and $\tau^a$ are isospin Pauli matrices.}
\label{gop}
\begin{tabular}{ll|l|l} \hline\hline
$\Lambda_\eta^\sigma$   & $k^{PC}$ & $H_n$ & $H_n(I=0,\,I=1)$ \\ \hline
$\Sigma_g^+$            & $0^{++}$ & $\mathbbm{1}$                 & $\left\{\bm{\bar{q}}T^a(\mathbbm{1},\,\bm{\tau})\bm{q}\right\}T^a$ \\
$\Sigma_u^-$            & $1^{+-}$ & ${\bf \hat{r}}\cdot{\bf B} $  & $\left\{\bm{\bar{q}}\,\left[(\hat{\bm{r}}\times\bm{\gamma})\cdot,\,\bm{\gamma}\right] T^a(\mathbbm{1},\,\bm{\tau})\bm{q}\right\}T^a$\\ 
$\Pi_u$                 & $1^{+-}$ & ${\bf \hat{r}}\times{\bf B}$  & $\left\{\bm{\bar{q}}\,\left[\hat{\bm{r}}\cdot\bm{\gamma},\,\bm{\gamma}\right] T^a(\mathbbm{1},\,\bm{\tau})\bm{q}\right\}T^a$\\
$\Sigma_g^{+\, \prime}$ & $1^{--}$ & ${\bf \hat{r}}\cdot{\bf E} $  & $\left\{\bm{\bar{q}}\, (\bm{\hat{r}}\cdot\bm{\gamma}) T^a(\mathbbm{1},\,\bm{\tau})\bm{q}\right\}T^a$\\
$\Pi_g$                 & $1^{--}$ & ${\bf \hat{r}}\times{\bf E} $ & $\left\{\bm{\bar{q}}\, (\bm{\hat{r}}\times\bm{\gamma})T^a(\mathbbm{1},\,\bm{\tau})\bm{q}\right\}T^a$\\ \hline\hline
\end{tabular}
\end{table}

\begin{figure}[ht]
\centering
\includegraphics[width=0.5\linewidth]{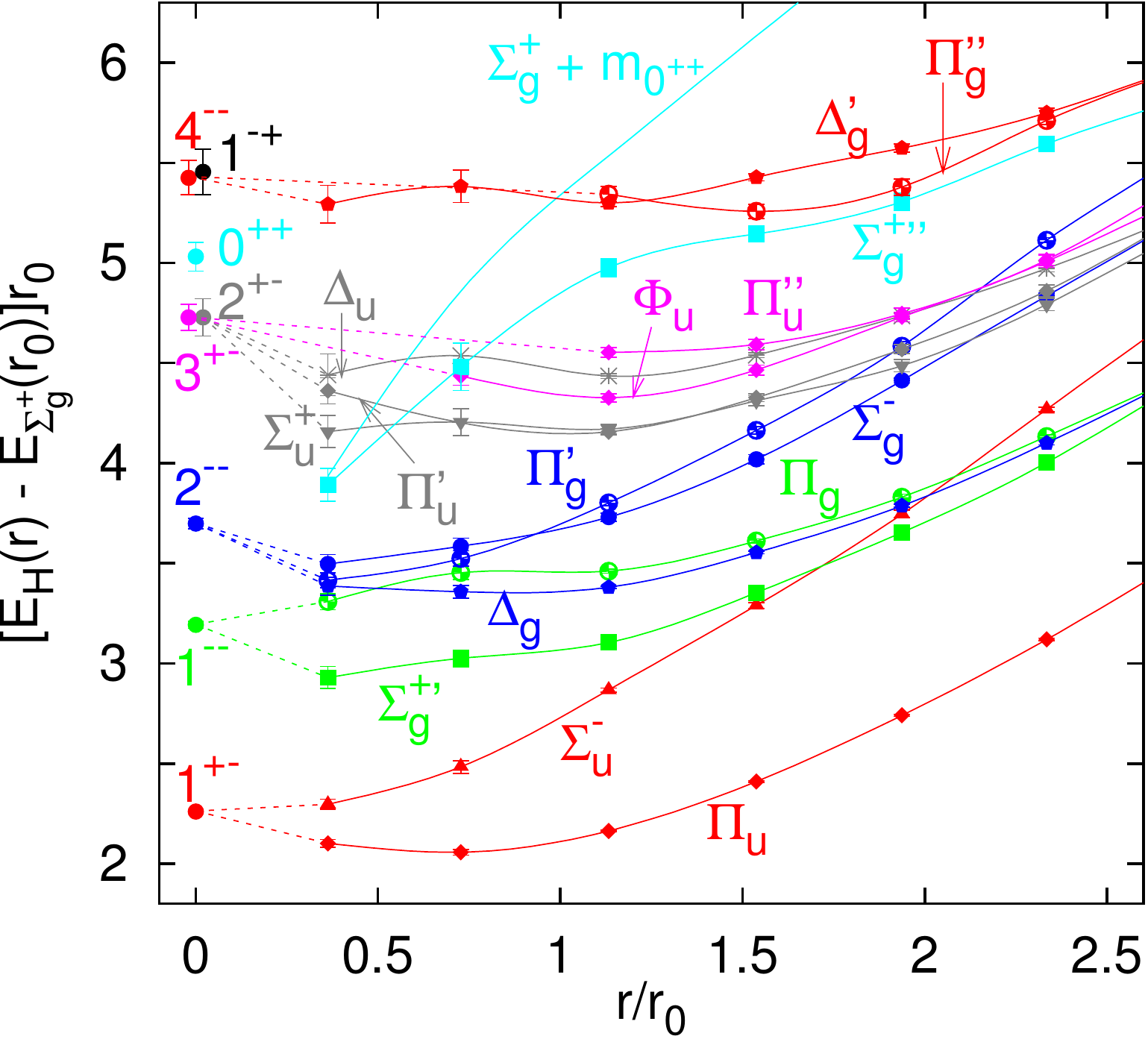}
\caption{The lowest hybrid static energies~\cite{Juge:2002br} and gluelump masses~\cite{Foster:1998wu,Bali:2003jq} in units of $r_0\approx 0.5$~fm. The absolute values have been fixed such that the ground state $\Sigma_g^+$ static energy (not displayed) is zero at $r_0$. The figure is taken from Ref.~\cite{Bali:2003jq}.}
\label{stelat}      
\end{figure}

The static energies associated to the gluonic operators in table~\ref{gop} have been computed using quenched lattice QCD in Refs~\cite{Juge:1997nc,Juge:2002br,Bali:2000vr,Capitani:2018rox}. In Fig.~\ref{stelat} we show the results from Ref.~\cite{Juge:2002br}. These static energies show interesting behaviors in the short- and long-distance limits. In the short distance the static energies form quasi-degenerate multiplets. This is a consequence of an enlargement of the symmetry of the system from $D_{\infty\,h}$ to $O(3)\otimes C$. The $k^{PC}$ representations each static state
tends to are displayed in the $r=0$ line in Fig.~\ref{stelat} as well as the second column of table~\ref{gop}. In this limit the gluonic excitations are commonly referred as gluelumps. In the long-distance limit the static energies become linear and correspond to the energy levels of a string~\cite{Nambu:1978bd,Polchinski:1991ax,PerezNadal:2008vm,Oncala:2017hop}, with the exited states corresponding to the number of phonons.

\begin{figure}[ht]
\begin{tabular}{cc}
\includegraphics[width=0.35\linewidth]{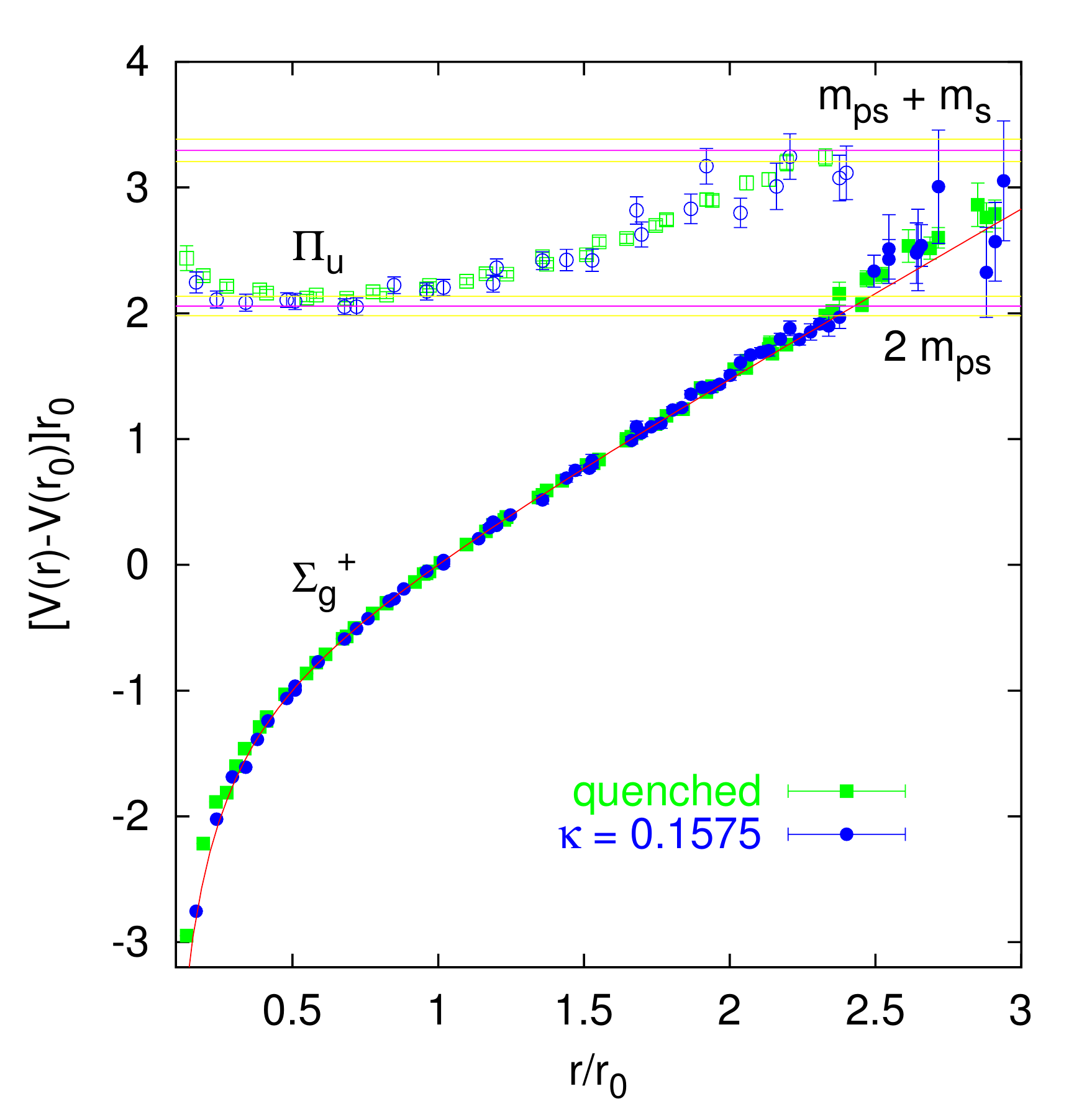} & \includegraphics[width=0.55\linewidth]{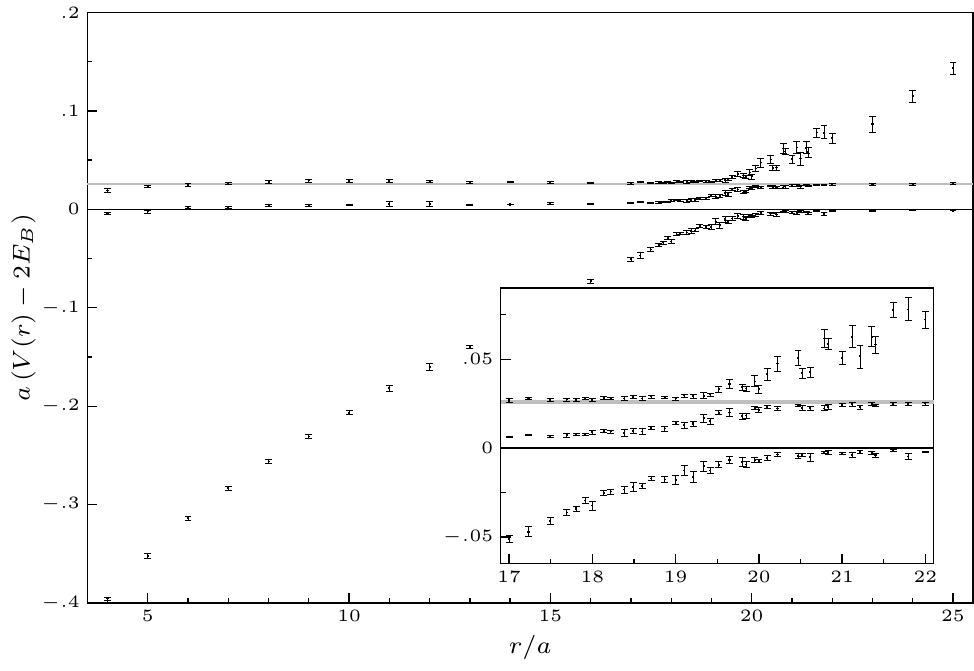} \\
(a) & (b) \\
\end{tabular}
\caption{(a): ground state and first excited static energies for $Q\bar{Q}$ from quenched (green) and unquenched (blue) lattice calculations of Ref.~\cite{Bali:2000vr}, figure taken from the same reference. (b): ground state $Q\bar{Q}$ static energy and the first two $Q\bar{q}$-$\bar{Q}q$ static energies from lattice QCD. Results and figure from Ref.~\cite{Bulava:2019iut}}
\label{lat1}
\end{figure}

So far we have only discussed the quenched case. An important question is how does the spectrum of static energies change when dynamical light-quarks are considered. Unfortunately, in this case we only have partial pieces of information from lattice QCD. First, let us consider the case in which the static state has isospin $I=0$. In Fig.~\ref{lat1}~(a) we shown the results from Ref.~\cite{Bali:2000vr} showing the two lowest laying static energies, which turn out no to have significant differences between the quenched and unquenched cases. No information is available for the further excited states but we expect a similar behavior to hold. However, when including light-quarks, we must also consider the appearance of new static energies associated to new kinds of states. The most important of these are the heavy-light meson pair thresholds, which appear as quasi-horizontal lines in the spectrum. When the meson pair static energies cross the ground state characteristic avoid crossings are observed~\cite{Bali:2005fu,Bulava:2019iut}, we illustrate this with Fig.~\ref{lat1}~(b) taken from Ref.~\cite{Bulava:2019iut}. Other possible states are $Q\bar{Q}$ states with light quark hadrons and charmed baryon pairs. Finally, there is the possibility that new static states associated to tetraquarks exist, however it is likely that they strongly mix with the gluonic
states studied in the quenched case and do not appear as separated energy levels in the unquenched one.

Let us finish the discussion on the static energies commenting the $I\neq 0$ case. From Refs.~\cite{Bali:2005fu,Alberti:2016dru} we know that the ground state for $I=1$ is approximately the $\Sigma_g^+$ $Q\bar{Q}$ static energy shifted up by one pion mass, and the first excited state corresponds to an $I=1$ heavy-light meson pair. It is likely, but not confirmed yet, that further excited states, analogous to the hybrid static energies~\cite{Brambilla:2008zz,Braaten:2014qka}, are present in the $I\neq 0$ spectrum of static energies. Such states would be unambiguously light quark in nature, since they would not mix with purely gluonic states, and would generate the charged tetraquark states which have been observed above open flavor thresholds in the spectrum of quarkonium-like states.

\section{Effective field theory for hybrids}

In this section we restrict ourselves to the case of quarkonium hybrids and build an EFT~\cite{Berwein:2015vca, Oncala:2017hop, Brambilla:2017uyf} that describes the heavy-quark bound states on the gluonic static energies discussed in the previous section. Non-relativistic bound states develop three well-separated scales $m_Q\gg m_Q v\gg m_Qv^2$, with $m_Q$ the heavy quark mass and $v$ the relative velocity. The heavy-quark--antiquark distance is of the order on the inverse momentum scale $r\sim 1/(m_Qv)$ and the binding energy is $E_b\sim m_Qv^2$. The bound states are small energy fluctuations around the minima of the static energies, thus the binding energy $E_b\sim\sqrt{\Lambda^3_{\rm QCD}/m}\ll \Lambda_{\rm QCD}$, and therefore the relative distance scales as $r\sim 1/(m_Qv)\lesssim 1/\Lambda_{\rm QCD}$. 

The hybrid EFT is obtained from NRQCD by integrating out the $\Lambda_{\rm QCD}$ modes. For simplicity and brevity in the following we are going to consider only the case of the static gluonic states associated to $k^{PC}=1^{+-}$ in table~\ref{gop} which correspond to the two lowest hybrid static energies. The Lagrangian reads

\begin{align}
L_{BO} &= \int d^3Rd^3r \sum_{\la\lap}\Psi^{\dagger}_{\lambda} \biggl\{i\partial_t - V_{\la\lap}(r)+\hat{\bm{r}}^{i\dag}_{\la}\frac{\bnabla^2_r}{m}\hat{\bm{r}}^i_{\lap}\biggr\}\Psi_{\lap}\,. \label{liso0}
\end{align}
The field $\Psi_{\lambda}$ has to be understood as depending on $t$, $\bm{r}$ and $\bm{R}$. The trace over spin indices is left implicit. The projectors are $\hat{r}^i_0=\hat{r}^i$ and $\hat{r}^i_\pm=\mp\left(\hat{\theta}^i\pm i\hat{\phi}^i\right)/\sqrt{2}$ where $\hat{\bm r} = (\sin\theta\cos\phi,\,\sin\theta\sin\phi\,,\cos\theta)$, $\hat{\bm \theta} =$ $(\cos\theta\cos\phi,$ $\,\cos\theta\sin\phi\,,-\sin\theta)$ 
and $\hat{\bm \phi} = (-\sin\phi,\,\cos\phi\,,0)$. The potential can be organized as an expansion in $1/m_Q$
\begin{align}
V_{\la\lap}(r)&=V^{(0)}_{\la}(r)\de_{\la\lap}+\frac{V^{(1)}_{\la\lap}(r)}{m_Q}+\frac{V^{(2)}_{\la\lap}(r)}{m_Q^2}+\dots
\end{align} 

It is interesting to study the matching in the short-distance regime $r\ll 1/\Lambda_{\rm QCD}$. In this case the scale associated to the relative heavy-quark momentum $mv\gg \Lambda_{\rm QCD}$ can be integrated out perturbatively leading to an EFT formally identical to weakly-coupled pNRQCD~\cite{Pineda:1997bj,Brambilla:1999xf}. One can then in turn integrate out the $\Lambda_{\rm QCD}$ modes and match weakly-coupled pNRQCD to the EFT for hybrids~\cite{Berwein:2015vca, Brambilla:2017uyf}. This procedure yields a short-distance description of the EFT potentials. The matching conditions from NRQCD to weakly-coupled pNRQCD to the hybrid and tetraquark EFT are

\begin{align}
g\mathcal{O}_n(t,\,\bm{r},\,\bm{R})\cong gZ^{1/2}_{o}(r) O^a(t,\,\bm{r},\,\bm{R})H^a_n(t,\,\bm{R})+\dots\cong Z^{1/2}_{o}(r)Z^{1/2}_{H_n}(\Lambda_{\rm QCD})\Psi_n(t,\,\bm{r},\,\bm{R})+\dots
\end{align} 
where $O^a$ is the color-octet heavy quark-antiquark field. $Z_{0}$ and $Z_{H_n}$ are normalization factors. The form of the potentials in the short distance can be generically described as follows: a sum of a perturbative part, which is typically nonanalytic in $r$ corresponding to the weakly-coupled pNRQCD potentials, and a nonperturbative part, which is a series in powers of $r$. The coefficients of the latter only depend on $\Lambda_{\rm QCD}$ and can be expressed in terms of gluonic and light-quark correlators. 

The static potential, $V^{(0)}_{\la}$, can be matched to the lattice NRQCD static energies and a short-distance weak-coupling pNRQCD description:

\begin{align}
E^{(0)}_{|\lambda|}(r)=V_o(r)+\Lambda_1+b_{|\lambda|}r^2+\dots=V^{(0)}_{\la}(r)\,,\label{semt}
\end{align} 
where $V_o(r)$ is the octet potential, $\Lambda_1$ is the (lowest lying) gluelump mass~\cite{Foster:1998wu}, which as well as $b_{|\lambda|}$ is a nonperturbative constant. The weakly-coupled pNRQCD description of the static energy in Eq.~\eqref{semt} reproduces the short-distance degeneracy we have observed in the lattice data in Fig.~\ref{stelat}, since up to next-to-leading order the potential is independent of the projection value $\lambda$.

\begin{figure}[ht!]
\centering
\includegraphics[width=0.85\linewidth]{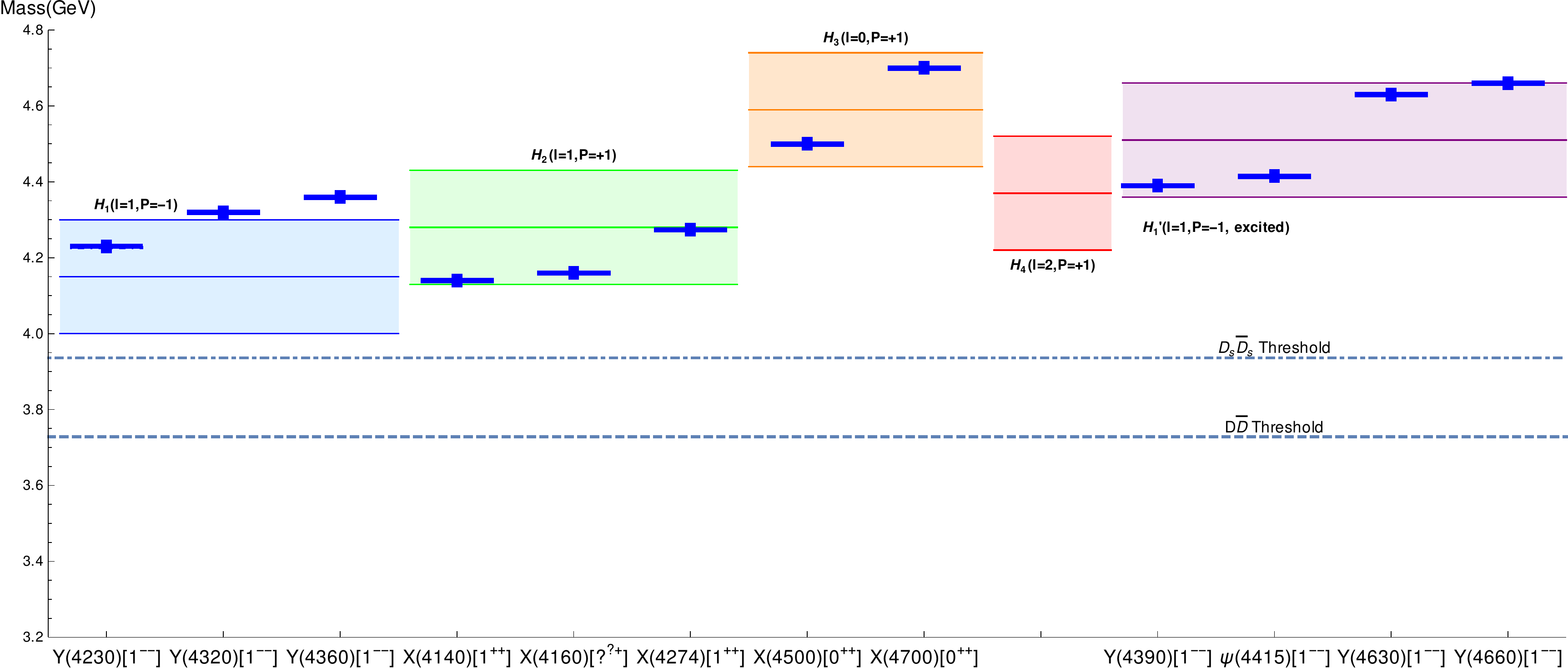}
\caption{Mass spectrum of neutral exotic charmonium states obtained by Ref.~\cite{Berwein:2015vca}. The experimental states of that have matching quantum numbers are plotted in solid blue lines. The spin-symmetry multiplets are those of Table~\ref{multi}, the prime indicates the first radially excited state. The multiplets have been plotted with error bands corresponding to a gluelump mass uncertainty of $0.15$~GeV. Figure updated from Refs.~\cite{Berwein:2015vca,Brambilla:2017uyf}.}
\label{cspect}      
\end{figure}

The kinetic operator in Eq.~\eqref{liso0} contains off-diagonal terms in $\la$-$\lap$ that mix the states associated to different projections. This off-diagonal matrix elements are the analog of the nonadiabatic coupling in molecular physics, and it is the leading correction that couples the dynamics of the heavy and light degrees of freedom. Unlike in diatomic molecules, the nonadiabatic coupling is not subleading in quarkonium hybrids due to the short-distance degeneracy of the potentials associated to the mixed states. The spectrum for quarkonium hybrids obtained by solving the coupled Schr\"odinger equations with the static potentials extracted from the lattice static energies was obtained in Refs.~\cite{Berwein:2015vca,Oncala:2017hop}. The results of Ref.~\cite{Berwein:2015vca} for the charmonium spectrum are shown in Fig.~\ref{cspect}  in bands representing the central value prediction and the uncertainty together with observations of neutral exotic charmonium. Each band corresponds to a spin symmetry multiplet of table~\ref{multi}. In several cases there are multiple experimental candidates to a given hybrid charmonium state. The spectrum for bottomonium hybrids can be found in Ref.~\cite{Berwein:2015vca}. In this sector the experimental states $\Upsilon(10750)$~\cite{Abdesselam:2019gth} and $\Upsilon(11020)$ are candidates for the ground and first radial exited states of the $H_1$ multiplet.

\begin{table}[ht!]
\begin{tabular}{c|c|c|c}\hline\hline 
        & $l$ & $J^{PC}\{s=0\,,s=1\}$       &   \\ \hline
  $H_1$ & $1$ & $\{1^{--},(0,1,2)^{-+}\}$ & $\Sigma_u^-$, $\Pi_u$ \\
  $H_2$ & $1$ & $\{1^{++},(0,1,2)^{+-}\}$ & $\Pi_u$ \\
  $H_3$ & $0$ & $\{0^{++},1^{+-}\}$       & $\Sigma_u^-$ \\
  $H_4$ & $2$ & $\{2^{++},(1,2,3)^{+-}\}$ & $\Sigma_u^-$, $\Pi_u$ \\ \hline\hline 
\end{tabular}
\caption{Spin-symmetry multiplets. Each multiplet corresponds to an angular eigenvalue momentum $l$, note that the angular momentum operator corresponds to the sum of the angular momentum of the heavy-quarks and the gluonic degrees of freedom. For each angular momentum $l>0$ the Sch\"odinger equations can be decoupled into two sets of equations for positive and negative parity, one of them still mixing the $\Sigma_u^-$ and $\Pi_u$ static states while the other is purely $\Pi_u$. The $J^{PC}$ of the hybrid quarkonium states are obtained combining the angular momentum state with the spin state of the heavy-quarks.}
\label{multi}
\end{table}

\section{Spin-dependent terms for quarkonium hybrids}

For the lowest-lying hybrid excitations of $k^{PC}=1^{+-}$, the spin-dependent potentials take the form 

\begin{align}
V_{\la\lap\,{\rm SD}}^{(1)}(\bm{r}) =& V_{{\rm SK}}(r)\left(\hat{r}^{i\dag}_{\la}\bm{K}^{ij}\hat{r}^j_{1\lap}\right)\cdot\bm{S}+ V_{{\rm SK}b}(r)\left[\left(\bm{r}\cdot \hat{r}^{\dag}_{\la}\right)\left(r^i\bm{K}^{ij}\hat{r}^j_{\lap}\right)\cdot\bm{S}+\left(r^i\bm{K}^{ij}\hat{r}^{j\dag }_{\la}\right)\cdot\bm{S} \left(\bm{r}\cdot \hat{r}_{\lap}\right)\right]\,,\label{sdm2}\\
V_{\la\lap\,{\rm SD}}^{(2)}(\bm{r}) =& V_{{\rm SL}a}(r)\left(\hat{r}^{i\dag}_{\la}\bm{L}_{Q\bar{Q}}\hat{r}^i_{\lap}\right)\cdot\bm{S}+V_{{\rm S}^2}(r)\bm{S}^2\de_{\la\lap}+V_{{\rm S}_{12}a}(r)S_{12}\de_{\la\lap}+ V_{{\rm SL}b}(r)\hat{r}^{i\dag}_{\la}\left(L_{Q\bar{Q}}^iS^j+S^iL_{Q\bar{Q}}^j\right)\hat{r}^{j}_{\lap}\nn\\
&+ V_{{\rm S}_{12}b}(r)\hat{r}^{i\dag}_{\la}\hat{r}^j_{\lap}\left(S^i_1S^j_2+S^i_2S^j_1\right)\,,\label{sdm3}
\end{align}
where $\bm{L}_{Q\bar{Q}}$ is the orbital angular momentum of the heavy-quark-antiquark pair, $\bm{S}_1$ and $\bm{S}_2$ are the spin vectors of the heavy quark and heavy antiquark, respectively, $\bm{S}=\bm{S}_1+\bm{S}_2$ and ${S}_{12}=12(\bm{S}_1\cdot\hat{\bm{r}})(\bm{S}_2\cdot\hat{\bm{r}})-4\bm{S}_1\cdot\bm{S}_2$. $\left({K}^{ij}\right)^k=i\epsilon^{ikj}$ is the angular momentum operator for the spin-1 representation. The first three operators on Eq.~\eqref{sdm3} are the standard spin-orbit, total spin squared and tensor spin operators that also appear in standard quarkonium, the rest of operators are unique to hybrid quarkonium. Specially noteworthy are the operators in Eq.~\eqref{sdm2} suppressed only by one power of the heavy quark mass. Since in standard quarkonium the leading spin-dependent operators appear at $1/m^2_Q$ order, it is to be expected that hyper-fine splitting will be more important in quarkonium hybrids than standard quarkonium. Additionally two new operators, in the second line of Eq.~\eqref{sdm3}, appear also at $1/m^2_Q$ order. Standard time-independent perturbation theory can be used to compute the mass shifts in the quarkonium hybrid spectrum produced by the spin-dependent operators.

In the short-distance limit the matching coefficients of the spin-dependent potentials can be determined matching weakly-coupled pNRQCD to the hybrid EFT. The matching coefficients can be generically characterized as the sum of a perturbative contribution and a nonperturbative one. The perturbative contribution corresponds to the spin-dependent octet potentials and only appears in the operators analogous to those of standard quarkonium. The nonperturbative contributions can be written as a polynomial series in $r^2$ with coefficients encoding the nonperturbative dynamics of the gluon fields. For example, the leading order operator matching coefficient $V_{{\rm SK}}$ is purely nonperturbative, since it has no standard quarkonium analog and can be written as

\begin{align}
V_{SK}&=V^{np\,(0)}_{SK}+V^{np\,(1)}_{SK}r^2+\dots\\
V^{np(0)}_{SK}&=\frac{c_F}{2}\lim_{T\to\infty}\frac{ie^{i\Lambda_1 T}}{T}
\int^{T/2}_{-T/2}dt\, \epsilon^{ijk}h^{bcd}\,\langle 0|\bm{B}^{ia}(T/2) \phi^{ab}(T/2,t)g \bm{B}^{jc}(t)\phi^{de}(T/2,t) \bm{B}^{ke}(-T/2)|0\rangle\,.\label{matchsk}
\end{align}

\begin{figure}[ht]
\centering
\includegraphics[width=0.45\linewidth]{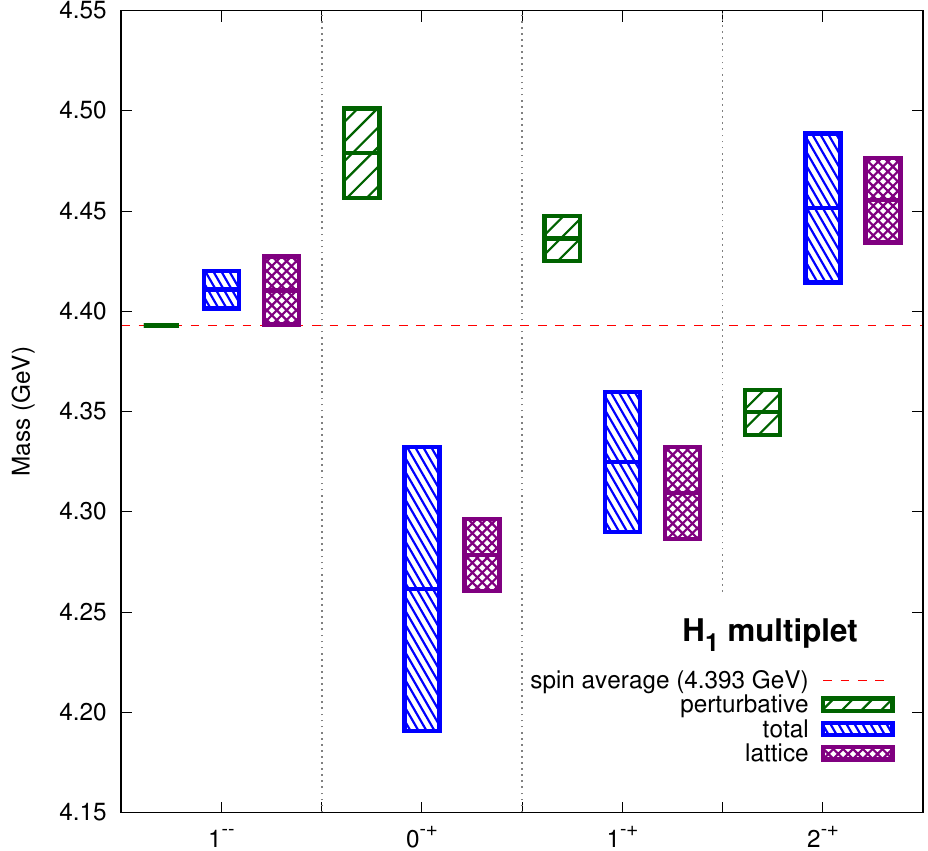}
\includegraphics[width=0.45\linewidth]{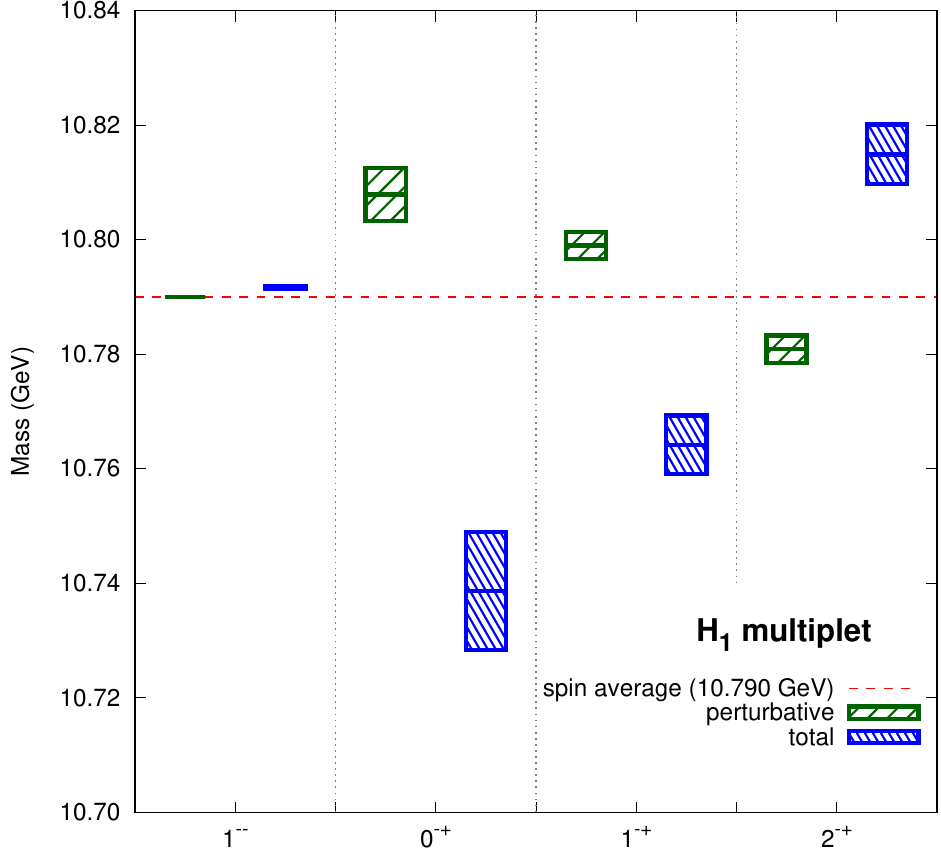}
\caption{Spectrum of the lowest-lying charmonium (left) and bottomonium (right) hybrid multiplet. The lattice results for the charmonium sector from Ref.~\cite{Cheung:2016bym} are the most right (purple) boxes for each quantum number. The perturbative contributions to the spin-dependent operators in Eq.~\eqref{sdm3} added to the spin average of the lattice results (red dashed lines) are the most left (green) boxes. The central (blue) boxes for each quantum number are the full results from the spin-dependent operators of Eqs.~\eqref{sdm2} and \eqref{sdm3} including perturbative and nonperturbative contributions. The height of the boxes indicates the uncertainty.}
\label{spspl}     
\end{figure}

It is interesting to note that all the heavy quark dependence of the matching coefficients can be factorized of the nonperturbative gluon correlators and its encoded in NRQCD matching coefficients such as $c_F$ in Eq.~\eqref{matchsk}. These nonperturbative gluon correlators can be obtained by fitting the spin splittings to the lattice determinations of the charmonium hybrid spectrum. Once the nonperturbative part of the matching coefficients is determined, the spin contributions in the bottomonium hybrid sector can be predicted~\cite{Brambilla:2018pyn}. In figure~\ref{spspl} we show the results for the lowest-lying hybrid multiplet.

\section{Hadronic transitions in quarkonium}

Hadronic transitions in quarkonium have been studied since the late 70's. The traditional approach has been to parametrize the transition amplitudes using chiral symmetry (current algebra in the early works and later on with EFT's) and to input of information on the bound state dynamics to partially or fully determine the free parameters left. To carry out the latter part of the program, it was noted that, in the context of the multipole expansion, hadronic transitions can be thought as  a two step process. First, a gluon is emitted by the initial quarkonium states and the singlet quarkonium state becomes a color-octet. This octet state propagates for a brief period of time and emits a second gluon transforming into the final singlet quarkonium state. In a longer time scale, dictated by $\Lambda_{\rm QCD}$, the two gluons hadronize into the final state pions. The multipole expansion allows for the derivation of selection rules and to relate transitions between different partial waves, however by itself does not allow for the full determination of the transition amplitude since it is still in terms of a nonlocal correlator involving the octet propagator and glue operators. The final step in the traditional approach, was proposed by Voloshin and consists in an operator product expansion of the octet propagator, which results in the amplitude being written in terms of local glue operators, for this reason this is often referred as the twist expansion of the transition amplitude. The hadronization of the local glue operators into the final pion states (at least the leading order term) can be determined making use of the scale~\cite{Voloshin:1980zf,Novikov:1980fa} and axial~\cite{Gross:1979ur,Novikov:1979uy} anomalies for the cases of two and one pions respectively. Nevertheless, the use of the twist expansion is only well justified when the energy of the final state pions is smaller than the typical binding energy of the quarkonium and therefore not valid for transitions between quarkonium states with different principal quantum number~\cite{Luty:1993xf}.

Our proposal is to study these transitions using an EFT approach, based on a hadronic version of weakly-coupled pNRQCD~\cite{Pineda:1997bj,Brambilla:1999xf}. The hadronic EFT is built in terms of the singlet (standard heavy quarkonium), hybrid and pion fields, organized according the $1/m_Q$ and multipole expansions as well as chiral symmetry. We also organize the computation according to the large $N_c$ expansion, which allows us to argue that $B/D$ mesons and possible tetraquarks contributions are, \textit{a priori}, subleading. In this EFT the intermediate octet states in the hadronic transitions correspond to the hybrid quarkonium spectrum and the amplitude can be written in terms of local glue operators which can be hadronized with the scale and axial anomalies leading to formally similar results as the ones using twist expansion. In Ref.~\cite{Pineda:2019mhw} we have studied $Q\bar{Q}(2S) \to Q\bar{Q}(1S)\pi\pi$, $Q\bar{Q}(2S) \to Q\bar{Q}(1S)\pi^0(\eta)$ and $Q\bar{Q}(2S) \to Q\bar{Q}(1P)\pi^0$ transitions. In the following we review the first case.

The hadronic version of the pNRQCD Lagrangian  reads 

\begin{align}
&L_{pNRQCD}^{\rm had} = \int d^3Rd^3r \,\Bigl[S^{\dag}\left(i\partial_t-V_s(r)+\frac{\bnabla^2_r}{m_Q}\right)\S+\sum_{\la\lap}\Psi^{\dagger}_{\la} \biggl\{(i\partial_t - (V_o^{(0)}+\Lambda_{k}))\delta_{\la\lap}+\hat{\bm{r}}^{i\dag}_{\la}\frac{\bnabla^2_r}{m_Q}\hat{\bm{r}}^i_{\lap}\biggr\}\Psi_{\lap} \nn\\
&+\frac{F^2}{4}\left(\langle u_{\mu}u^{\mu}\rangle+\langle\chi_+\rangle\right)+\left(\bm{r}\cdot \hat{\bm{r}}_{\la}S^{\dag}\Psi_{\la}+\text{h.c}\right)\left(t^{(r1^{--})}+t^{(r1^{--})}_{d0}F^2\langle u_0u_0\rangle+t^{(r1^{--})}_{di}F^2\langle u_i u^i\rangle+t^{(r1^{--})}_m F^2\langle\chi_+\rangle\right)\Bigr]\,.
\label{bolag2}
\end{align}

The fields $S$ and $\Psi_{\lambda}$ should be understood as depending on $t$, $\bm{r}$ and $\bm{R}$ and correspond to the standard and hybrid quarkonium. In this case the hybrid quarkonia we consider are the states with $k^{PC}=1^{--}$ unlike those in Eq.~\eqref{liso0}. The unitary matrix $u=exp(i\bm{\pi}\cdot\bm{\lambda}/(2F))$ contains the pion fields, which depend on $t$ and $\bm{R}$, F is the pion decay constant and

\begin{align}
u_{\mu}=&i\left(u^{\dag}(\partial_{\mu}-i r_{\mu}))u-u(\partial_{\mu}-i l_{\mu})u^{\dag}\right)\,, \\
\chi_{\pm}=&u^{\dag}\chi u^{\dag}\pm u \chi^{\dag} u\,,
\end{align}
where $\chi =2B {\rm diag}(\hat{m},\hat{m},m_s)$, with $B$ being related to the vacuum quark condensate. In the isospin limit, the pion mass is $m_{\pi}^2 = 2 B\hat{m}$. $\langle A \rangle$ stands for the trace of $A$ in the isospin index and the trace over spin indices is left implicit.

The singlet-hybrid mixing term in the hadronic theory is given by

\begin{align}
Z^{1/2}_E\langle 0|\Psi_{\la}(t,\,\bm{r}',\,\bm{R}')S^{\dag}(0,\,\bm{r},\,\bm{R})|0 \rangle_{\rm amp.}=i \bm{r}\cdot \hat{\bm{r}}^{\dag}_{\la}Z^{1/2}_Et^{(r1^{--})}\,,\label{s1mmmixa}
\end{align}
which according to the matching condition in Eq.~\eqref{intop} is matched to the following correlator of pNRQCD in term of quarks and gluons:

\begin{align}
&g\langle 0|\hat{\bm{r}}^{\dag}_{\la}\cdot \bm{E}^{a\dag}(t,\bm{R}')O^{a}\left(t,\,\bm{r}',\,\bm{R}'\right)S^{\dag}(0,\,\bm{r},\,\bm{R})|0\rangle_{\rm amp.}=i\sqrt{\frac{T_F}{N_c}}\hat{\bm{r}}^{\dag i}_{\la}\bm{r}^j \langle 0|g\bm{E}^{a i\dag}(\bm{R})g\bm{E}^{aj}(\bm{R})|0\rangle
\,,\label{s1mmmixb}
\end{align}
where \textit{amp.} signals that only amputated contributions are considered (overall $\delta({\bf r}'-{\bf r})$ are also factored out). The normalization of the $\Psi_{\la}$ field implies

\begin{align}
 \langle 0|g\bm{E}^{a i\dag}(\bm{R})g\bm{E}^{aj}(\bm{R})|0\rangle=Z_E\delta^{ij}+\dots\label{normee}
\end{align}

Putting Eqs.\eqref{s1mmmixa}-\eqref{normee} together we arrive at

\begin{align}
t^{(r1^{--})}=\sqrt{\frac{T_F Z_E}{N_c}}\,.
\end{align}

The operators with an even number of pions in Eq.~\eqref{bolag2} are matched in a similar way. In the hadronic EFT

\begin{align}
& Z^{1/2}_E \int d^4x_+d^4x_-e^{i p_+\cdot x_+}e^{ip_-\cdot x_-}\langle 0|\pi^+(x_+)\pi^-(x_-)S(t,\,\bm{r},\,\bm{R})\Psi^{\dag}_{\la}(0,\,\bm{r},\,\bm{R})|0\rangle_{\rm amp.}\nn\\
&=i4\,Z^{1/2}_E\bm{r}\cdot \hat{\bm{r}}_{\la}\left(-t_{d0}^{(r1^{--})}p^0_+p^0_-+t_{di}^{(r1^{--})}\bm{p}_+\cdot\bm{p}_--t_{m}^{(r1^{--})}m^2_{\pi}\right)\label{2pma}\,,
\end{align}
and the corresponding correlator in pNRQCD reads

\begin{align}
&\langle \pi^+(p_+)\pi^-(p_-)|S(t,\,\bm{r},\,\bm{R})\hat{\bm{r}}_{\la}\cdot \bm{E}^a(0,\,\bm{R}) O^{a\dag}(0,\,\bm{r},\,\bm{R})|0\rangle_{\rm amp.}=ig\sqrt{\frac{T_F}{N_c}}\hat{\bm{r}}_{\la}\cdot\bm{r}\langle \pi^+(p_+)\pi^-(p_-)|\bm{E}(\bm{R})\cdot \bm{E}(\bm{R})|0\rangle\nn\\
&=\frac{i}{3}\sqrt{\frac{T_F}{N_c}}\hat{\bm{r}}_{\la}\cdot\bm{r}\frac{8\pi^2}{\beta_0}\left(\left(2-\frac{9}{2}\kappa\right)p^0_+p^0_--\left(2+\frac{3}{2}\kappa\right)\bm{p}_+\cdot\bm{p}_-+3m^2_{\pi}\right)\,,\label{2pmb}
\end{align}
where in the last step we hadronize the gluonic matrix element using the anomaly relation of the energy-momentum tensor of QCD \cite{Voloshin:2007dx,Novikov:1980fa} which determines the matrix element up to the free parameter $\kappa$.

Comparing Eq.~\eqref{2pma} and Eq.~\eqref{2pmb} we obtain

\begin{align}
&t_{d0}^{(r1^{--})}=-\frac{2\pi^2}{3\beta_0}\sqrt{\frac{T_F}{N_cZ_E}}\left(2-\frac{9}{2}\kappa\right)\,,\,t_{di}^{(r1^{--})}=-\frac{2\pi^2}{3\beta_0}\sqrt{\frac{T_F}{N_cZ_E}}\left(2+\frac{3}{2}\kappa\right)\,,\,t_{m}^{(r1^{--})}=-\frac{2\pi^2}{\beta_0}\sqrt{\frac{T_F}{N_cZ_E}}\,.
\end{align}

For comparison purposes let us write the most general transition amplitude for two pion transitions according to chiral and heavy quark spin symmetry up to ${\cal O}(p^2)$

\begin{align}
\mathcal{A}_{\chi}&=-a_1 p^0_+p^0_-+a_2 \bm{p}_+\cdot \bm{p}_--a_3m^2_{\pi}\,,
\label{2pgp1}
\end{align}

Using the hadronic pNRQCD Lagrangian of Eq.~\eqref{bolag2} the free parameters of Eq.~\eqref{2pgp1} take the form

\begin{align}
&a_1=-\frac{8\pi^2T_F}{3\beta_0 N_c}\beta^{(12)}_{r,Q}\left(2-\frac{9}{2}\kappa\right)\,,\,a_2=-\frac{8\pi^2T_F}{3\beta_0 N_c}\beta^{(12)}_{r,Q}\left(2+\frac{3}{2}\kappa\right)\,,\,a_3=-\frac{8\pi^2T_F}{\beta_0 N_c}\beta^{(12)}_{r,Q}\,,\label{c21a3}
\end{align}
with $\beta^{(12)}_{r,Q}$ the sum over the intermediate $k^{PC}=1^{--}$ hybrid states

\begin{align}
&\beta^{(n^{\prime}n)}_{r,Q}=\sum_m \langle S_{n^{\prime}} | \hat{\bm{r}}^{\dag}_{\la}\cdot\bm{r}|\Psi_m\rangle\left(\frac{1}{m_{n}-m_m}+\frac{1}{m_{n'}-m_m}\right)\langle\Psi_m|\hat{\bm{r}}^{\dag}_{\la}\cdot\bm{r}|S_n\rangle \,.\label{betarc21}
\end{align}

It is remarkable that the normalization factors $Z_E$ cancels out, which allows us to completely evaluate the transition amplitude except for the parameter $\kappa$.

\begin{figure}[ht]
\includegraphics[width=.6\textwidth]{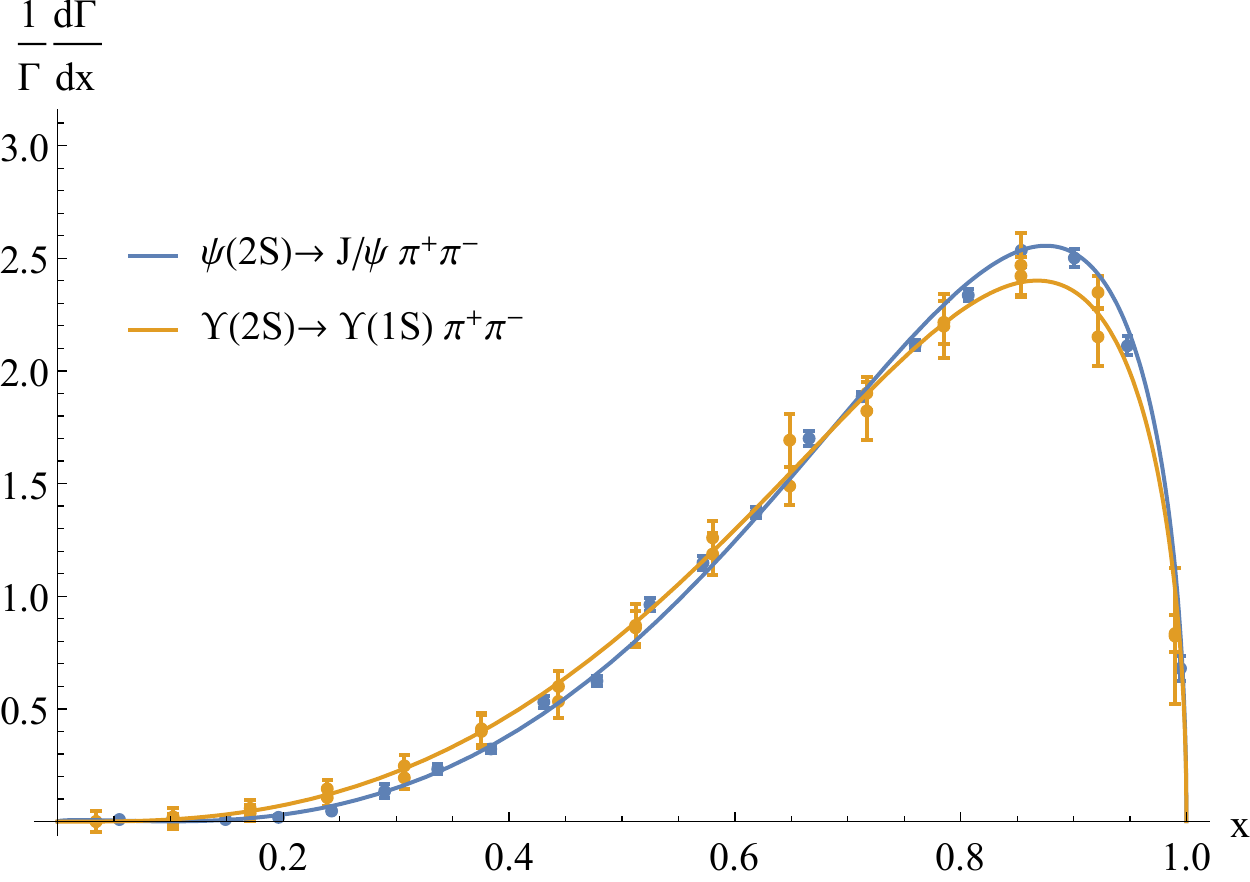}\\
\caption{Plot of the normalized differential decay width spectrum. The dots are the experimental data for $\psi(2S)\to J/\Psi \pi^+\pi^-$~\cite{Aaboud:2016vzw} and $\Upsilon(2S)\to \Upsilon(1S) \pi^+\pi^-$~\cite{Alexander:1998dq} in blue and yellow respectively. In the same color scheme the continuous lines are the fits of the theoretical expression obtained from the amplitude computed with the hadronic pNRQCD Lagrangian. The variable $x$ is defined as $x=\frac{m_{\pi\pi}-2m_{\pi}}{m_{2S}-m_{1S}-2m_{\pi}}$.}
\label{fig:kappa}
\end{figure}

We can now compare our results with experimental data. First let us look at the normalized decay width spectrum with respect to the dipion invariant mass. This normalized distribution is independent of the unknown normalization of the experimental spectrum making it a convenient observable to compare with, moreover the theoretical expression of the normalized decay width spectrum is independent of $\beta^{(n^{\prime}n)}_{r,Q}$ and only depends on $\kappa$. Therefore, we can obtain the value of $\kappa$ by fitting the normalized decay width spectrum. The results are shown in Fig.~\ref{fig:kappa}.

The values for $\kappa$ obtained from the fits are as follows
\begin{align}
\kappa_c=0.277\pm 0.015\,,\quad \kappa_b=0.229\pm 0.016 \,,\quad \kappa_{\rm joint}=0.247^{+0.014}_{-0.013}\,.\label{kappajointrel}
\end{align}
We will take the value of the joint fit to both sets of experimental data, $\kappa_{\rm joint}$, in our numerical evaluations of the total widths. To estimate the uncertainty associated to subleading effects we take the difference of the value in Eq.~\eqref{kappajointrel}, obtained with phase space factors computed with relativistic kinematics, with the value obtained using nonrelativistic kinematics. We then combine them in quadrature with the statistical error quoted in Eq.~\eqref{kappajointrel} and give $\kappa=0.247(20)$.

We compute $\beta^{(12)}_{r,Q}$ with the $k=1^{--}$ and $m^3S_1$ hybrid intermediate states $m=1,..,4$ (we observe that the effect of introducing 3 or 4 hybrid states is comparatively small compared with other uncertainties). The hybrid masses and wave functions are obtained using the techniques developed in Ref.~\cite{Berwein:2015vca}. We obtain the following values for the decay widths
\begin{align}
\label{Gammacpipi}
&\Gamma_{\psi(2S)\to J/\psi\pi^+\pi^-}=46.2(_{+15.7}^{-10.3})_{\Lambda_1}(_{+3.4}^{-3.2})_{\kappa}(\pm 21.2)_{\rm s.p.}\,{\rm keV}\,, \quad \Gamma^{\rm exp}=102.1(2.9) \,{\rm keV}\,,\\
\label{Gammabpipi}
&\Gamma_{\Upsilon(2S)\to\Upsilon(1S)\pi^+\pi^-}=3.08(_{+0.81}^{-0.58})_{\Lambda_1}(_{+0.22}^{-0.23})_{\kappa}(\pm 1.49)_{\rm s.p.}\,{\rm keV}\,, \quad \Gamma^{\rm exp}=5.71(48) \,{\rm keV}\,.
\end{align}
The uncertainties are labeled according to the source, with $\Lambda_1$ the lowest lying gluelump mass, and s.p. the different parametrization for the singlet and hybrid static potentials. Our numbers differ from the experimental ones by about a factor 2. One should keep in mind however that our estimates suffer from large uncertainties. We find a significant dependence on variations of the wave function of the hybrid and singlet state. The error generated by the uncertainty on the energy difference between singlet and hybrid states is somewhat smaller. These error estimates are of the right magnitude, though not large enough, to completely account for the difference with experiment. One should keep in mind however that, besides those errors already estimated, one error that has not been incorporated in this analysis is due to the uncertainties associated to the hadronization of the local operator such as ${\cal O}(\alpha_s)$ corrections to the beta function. 

It is interesting to consider the ratio of decay widths, where the uncertainties associated to the gluon hadronization cancel out, as well as a reduction on the impact of the other uncertainty sources 
\begin{align}
\label{Rbcpipi}
R_{bc,\pi\pi}\equiv R\left(\frac{\Upsilon(2S)\to \Upsilon(1S) \pi^+\pi^-}{\psi(2S)\to J/\psi \pi^+\pi^-}\right) = & 6.65(^{+0.30}_{-0.38})_{\Lambda_1}(\mp 0.02)_{\kappa}(\pm 0.31)_{\rm s.p.}\times 10^{-2}
\,,\\\nn
R_{bc,\pi\pi}^{\rm exp}= & 5.59(0.50)\times 10^{-2}\,.
\end{align}
This observable can be considered a rough measure of $\beta_{r,b}^{(21)}/\beta_{r,c}^{(21)}$. The agreement with experiment is remarkable: below 20\%, and could be accounted for by the quoted errors. 

\section{Conclusion}

We have reviewed recent developments in the EFT approach for exotic quarkonium. In particular, we have reported the calculation of the hybrid quarkonium spectrum including spin-dependent contributions, which owning to their appearance at $1/m_Q$ suppression instead of $1/m^2_Q$ as in standard quarkonium are expected to be of enhanced importance. Finally we have shown how the hybrid quarkonium spectrum can be used to compute the intermediate states in quarkonium hadronic transitions in a similar EFT framework.

\begin{acknowledgments}
This work was supported in part by the Spanish grants FPA2017-86989-P and SEV-2016-0588 from the Ministerio de Ciencia, Innovaci\'on y Universidades, and the grant 2017SGR1069 from the Generalitat de Catalunya. J.T.C acknowledges the financial support from the European Union's Horizon 2020 research and innovation programme under the Marie Sk\l{}odowska--Curie Grant Agreement No. 665919.
\end{acknowledgments}

\nocite{*}
\bibliographystyle{apsrev4-2}
\bibliography{bibprmenu}% Produces the bibliography via BibTeX.

\end{document}